\documentstyle[prd,aps,floats]{revtex}
\begin{document}
\draft
%
%
\input epsf
\renewcommand{\topfraction}{0.8}
\twocolumn[\hsize\textwidth\columnwidth\hsize\csname
@twocolumnfalse\endcsname
\title{Gamma photons from parametric resonance in neutron stars} 
\author{Juan Garc{\'\i}a-Bellido}
\address{Theoretical Physics, Imperial College, Blackett Laboratory, 
Prince Consort Road, London SW7 2BZ, U.K.}
\author{Alexander Kusenko}
\address{Department of Physics and Astronomy, University of California, Los
Angeles, CA 90095-1547, U.S.A.}
\date{December 28, 1998} 
\maketitle

\begin{abstract}

Shock waves in cold nuclear matter, {\it e.\,g.}, those induced by a 
collision of two neutron stars, can generate a large number of gamma
photons via parametric resonance.  We study the resonant production of
gamma rays inside a shocked neutron star and discuss the possible
astrophysical consequences of this phenomenon. 

\end{abstract}

\pacs{PACS numbers: 26.60.+c, 98.70.Rz  \hspace{1.0cm} 
IMPERIAL/TP/98-99/30, UCLA/99/TEP/5}

\vskip2pc]

\section{Introduction}

Neutron star collisions rank among the most energetic events expected
to take place in the Universe, which makes them a natural candidate
for the source of observed gamma-ray bursts (GRB)~\cite{Rees}. The
discovery of the afterglow~\cite{afterglow} associated with some of
the GRB and the isotropy of the GRB both support the hypothesis that
the GRB originate at cosmological distances with a redshift of order
one.  The total energy $\sim 10^{53}$~erg released when the two
neutron stars merge is sufficient to explain the GRB, although it is
not clear what fraction of that energy is converted to gamma rays.
Another puzzling feature of the GRB is their non-thermal spectrum.

Understanding the physics of the explosion that follows the collision of
two neutron stars is of great importance and is the subject of intense
studies~\cite{cns}.  The approach and the early stages of the interaction
of the two neutron stars are accompanied by powerful acoustic shock waves
that propagate through nuclear matter and eventually dissipate their energy
into heat.  During this period of time, the repetitive superconducting
phase transitions can take place in part of the star's volume due to the
sharp dependence of the proton energy gap on density.  The relaxation of
the proton condensate to the potential minimum can be accompanied by a
non-thermal resonant production of gamma rays~\cite{ak} in the MeV energy
range.  If the parametric resonance is efficient, the power transferred to
the coherent gamma quanta may exceed $10^{52}$~erg/s.  Eventually, the
nuclear matter is heated above the critical temperature $\sim 0.5$~MeV and
the production of the gamma rays comes to a halt.

The photons produced inside a neutron star cannot decay through
pair production $\gamma \rightarrow e^+e^-$ because the degenerate
electrons have the chemical potential in excess of 100~MeV, far
greater than the photon energy.  The decay into the electron-positron
pairs is, therefore, prohibited by the Pauli exclusion principle.  The
photons undergo a Compton scattering off the electrons near the Fermi
surface which is sufficiently strong to keep the photons from
escaping.   While such scattering (or comptonization) may change the
spectrum of the gamma-component, the number of photons remains the
same.  Since the extreme electron degeneracy is maintained even at the
highest temperatures achieved in the fireball~\cite{cns}, the
gamma-ray component is present in the nuclear matter at the time the
latter is dispersed by the explosion.  

The latter can have several consequences.  In particular, $\gamma$-quanta,
abundantly present in nuclear matter at the onset of the explosion,
can be released when the fireball erupts.  The detailed investigation
of this signal lies outside the scope of this paper.  However, it is
plausible that the gamma-rays emitted at that point would have a
non-thermal spectrum.  At later times, when the fireball reaches a
high temperature, other sources of gamma emission become dominant.  We
predict, therefore, a qualitative difference in the spectrum of
gamma-rays emitted during the first milliseconds of the collision.

In this paper we will not attempt to understand the emission of gamma-rays
from the fireball.  Instead, we will concentrate on the phenomenon of
resonant production of photons which transfers a fraction of the
gravitational energy into the gamma-quanta inside the neutron star. 
This process is of fundamental interest on its own and it may have
important consequences.  

\section{Basic idea} 
\begin{figure}[t]
\centering
\hspace*{-4mm}
\leavevmode\epsfysize=5.6cm \epsfbox{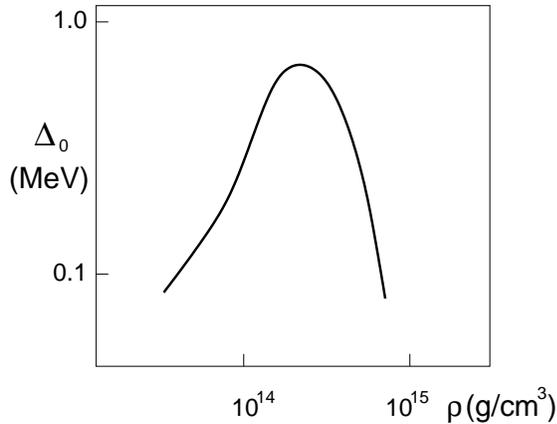}\\[3mm]
\caption[fig0]{\label{fig0} The energy gap as a
function of matter density inside a neutron
star~\cite{egap}.  Superconductivity occurs in some range of densities, 
$10^{14} < \rho < 6 \times 10^{14}$~g/cm$^3$.}
\end{figure}

%
Nuclear matter in the interior of a neutron star may be
superconducting. The existence of the superfluid proton condensate 
depends on several theoretical assumptions, some of which may be hard to
justify. The proper usage of the many-body techniques and the choice of
macroscopic degrees of freedom are by no means obvious. However, assuming
that the theoretical framework of Refs.~\cite{super,egap,chen}
is valid, one can elucidate some generic features of proton superfluidity.
The most important one for us is that the energy gap depends sharply
on density, as shown in Fig.~\ref{fig0}. 
An
acoustic shock wave passing through the star can produce significant changes
in the density and cause repetitive superconducting phase transitions
occurring with periods of order the acoustical time scale $\tau_a \sim
10^{-7}$ to $10^{-3}$~s~\cite{spectrum}.  The relaxation time of the proton
condensate is of order MeV$^{-1}$, or $10^{-20}~{\rm s} \ll \tau_a$, and
therefore the system quickly settles in the potential minimum after the
passage of each shock wave.

In the presence of a magnetic field the superconducting phase transition is
first-order~\cite{super} and occurs in two stages.  First, a bubble of the
superconducting phase nucleates and expands; then the proton condensate may
oscillate around the minimum.  If the magnetic field is
close to critical, $B\sim B_c$, the two minima in the potential shown in
Fig.~\ref{fig1} are nearly degenerate and no subsequent coherent motion of
the condensate $\phi$ takes place after the transition through bubble
nucleation.  However, for a smaller magnetic field, the so called ``escape
point'' $\phi_e$ is different from the global minimum $\phi_0$ and the
scalar condensate may oscillate around the minimum.

During the oscillations of the order parameter $\phi$, a photon has
effectively a time-dependent mass proportional to the value
of $\phi$.  This may, in some cases, signal a copious production of
gamma-quanta through parametric resonance~\cite{param}.  In this paper
we study the parametric resonance both analytically and numerically for
different for some sample values of nuclear density and magnetic field that
ranges from $10^{12}$ to $10^{15}$~G.

\section{Potential and equation of motion}

A natural framework for describing the
relaxation of the proton condensate after the phase transition is
time-dependent Ginzburg-Landau theory~\cite{tinkham,at}.  The applicability
of such description is limited to cases where the Joule heat loses are
small~\cite{ge}.  Otherwise, the interactions become essentially
non-local and a simple equation of motion for a scalar field seizes to
be valid.

Here we will assume the validity of the time-dependent Ginzburg-Landau
theory~\cite{at,ge} and will apply it to the superconducting proton
condensate inside a neutron star.  One should take this approach with
a grain of salt given the lack of empirical knowledge with respect to
the proton superconductivity in nuclear matter.  However, we hope that
-- crude an approximation this may be -- it may help identify the main
features of the resonant production of gamma-rays.

We write the equation of motion for the order parameter as 
\begin{equation}\label{eqm}
\ddot\phi + {8\varepsilon_F\over3c}\,\dot\phi -
{2\varepsilon_F\over3c\,m_*}\,\nabla^2\phi + U'(\phi) = 0\,,
\end{equation}
where $\varepsilon_F=p_F^2/2m_*$ is the Fermi energy of the proton
condensate and $c=(28\zeta(3)/3\pi^3)\,\varepsilon_F/T_c$ is a constant
characterizing the condensate. The friction term then has a magnitude
$8\varepsilon/3c \simeq 7.37 T_c \simeq 4.2 \Delta_0$, of order a few MeV,
which is comparable to the oscillating frequency $\omega$ of the condensate
around its minimum $\phi_0$, and therefore cannot be ignored. On the other
hand, the gradient term in Eq.~(\ref{eqm}) is negligible in our
case~\cite{ak}, and thus we are dealing with a (locally) homogeneous
condensate.

The effective potential for the order parameter $\phi$ can be
determined from the physical properties of the condensate. After the
acoustic wave has restored the symmetry, the energy density subsides
and a new superconducting phase transition takes place. The difference
in energy density associated with the proton condensate
is~\cite{ll}
\begin{equation}\label{du}
\Delta U(\phi_0) = {m_*p_F\over4\pi^2}\,\Delta_0^2\,,
\end{equation}
where $\Delta_0$ is the proton energy gap, and $m_*$ is the effective
proton mass in nuclear matter, which is somewhat lower than the bare proton
mass $m_p$.  The equilibrium value of the order parameter is given by
$\phi_0^2= n_p/2m_*$, where $n_p=Y_p \rho/m_p$ is the number density of
protons inside the neutron star, which make up a fraction $Y_p\approx0.03$
of all baryons. 

\begin{figure}[t]
\centering
\hspace*{-4mm}
\leavevmode\epsfysize=5.6cm \epsfbox{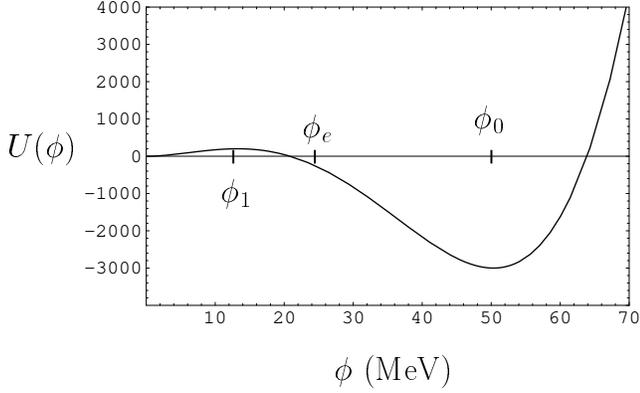}\\[3mm]
\caption[fig1]{\label{fig1} Effective potential for the proton
condensate, in units of MeV$^4$. The small bump at $\phi_1$ is
due to a magnetic field $B$.}
\end{figure}

The way the superconducting phase transition occurs depends very
strongly on the presence of a magnetic field $B$. Such a field creates
a barrier in the effective potential between the symmetric and
the superconducting phases. The height of the barrier is approximately
given by the total energy density in the magnetic field,
\begin{equation}\label{dB}
\Delta U(\phi_1) = {B^2\over4\pi} \simeq 200\,
\Big({B\over10^{15}~{\rm G}}\Big)^2\,{\rm MeV}^4\,.
\end{equation}
This barrier  makes the phase transition first-order, with
creation of bubbles of superconducting phase. After the bubble nucleation, 
the proton condensate oscillates and soon settles (due to friction) at its
equilibrium value $\phi_0$. We have shown in Fig.~\ref{fig1} the
effective potential $U(\phi)$ for a region inside the neutron
star with matter density $\rho=10^{14}$~g/cm$^3$ and magnetic field
$B=10^{15}$~G, which corresponds to $\Delta_0\simeq0.35$~MeV,
$\phi_0\simeq50.3$~MeV, $\Delta U(\phi_0)\simeq3000$~MeV$^4$ and
$\Delta U(\phi_1)\simeq200$~MeV$^4$.

\begin{figure}[t]
\centering
\hspace*{-4mm}
\leavevmode\epsfysize=5.6cm \epsfbox{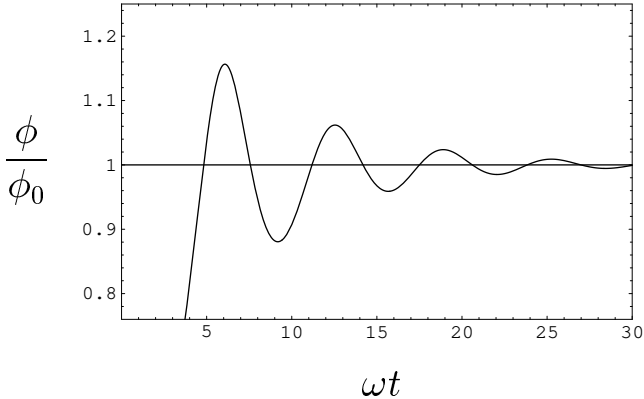}\\[3mm]
\caption[fig2]{\label{fig2} The oscillations of the order parameter
associated with the proton condensate around the minimum of the 
effective potential. Time is given in units of the effective frequency.
Due to the friction term, 
the amplitude of the proton condensate diminishes in a few oscillations.}
\end{figure}

Depending on the relative size of the friction term, the condensate settles
at $\phi_0$ after a few oscillations, or without oscillations at all.  We
have shown in Fig.~\ref{fig2} the oscillations of the order parameter
$\phi$ in the case of the parameters of Fig.~\ref{fig1}, which give a
friction coefficient of order $4.2\,\Delta_0 \simeq 0.31\,\omega$ and
$\omega\simeq 4.7$~MeV. This allows a few oscillations to take place, which
is crucial for the resonant production of photons, as we will see in the
next Section.

\section{Resonant photoproduction}

The proton condensate comprises Cooper pairs of charge $2e$,
which couple to the electromagnetic field through the term 
$(2e)^2\phi^2\,A_\mu A^\mu$ in the unitary gauge, in which the scalar
field $\phi$ is real. As the condensate oscillates around $\phi_0$, it
induces an effective periodic mass, which in some cases may stimulate
parametric resonant production of photons~\cite{param}. The
mechanism is similar to that discussed in connection with axion
clumps~\cite{tkachev}, as well as {\em pre}heating after
inflation~\cite{KLS}, where explosive production of bosons may occur
under special circumstances~\cite{GBL}.

We will write the gauge field as $A_\mu(x) = \chi(x)\,e_\mu(x)$, where
$e_\mu(x)$ is a polarization vector and $\chi(x)$ can be expanded
in Fourier modes, $\chi_k(t)$, that satisfy the evolution equation
\begin{equation}\label{chi}
\ddot\chi_k + \Big(k^2 + 2(2e)^2\phi^2(t)\Big)\chi_k(t) = 0\,,
\end{equation}
with an effective mass proportional to the condensate 
\begin{equation}
\phi(t) \simeq \phi_0\Big(1 + \Phi \exp(-\epsilon\,\omega t/2) 
\sin\omega t\Big)\,,
\end{equation}
where $\Phi=\phi_e/\phi_0$ and $\epsilon\simeq4.2\Delta_0/\omega$ is
the decay constant for the condensate oscillations.
Equation~(\ref{chi}) can be written as a Mathieu equation~\cite{math}
with coefficients ($z=\omega t/2$)
\begin{eqnarray}
A_k &=& {4k^2\over\omega^2} + 4q_0\,,\\
q(z) &=& 4q_0 \Phi \exp(-\epsilon z)\,,\\
q_0 &=& 8e^2\,{\phi_0^2\over\omega^2} =
32\pi^2\,\alpha_{\rm em}\,{\phi_0^2\over\omega^2}\,.
\end{eqnarray}
Note that the effective parameter $q(z)$ that determines the strength
of the resonance decreases exponentially with time. If $\epsilon$ is
too large, the parametric resonance is weak, and this mechanism is
inefficient. On the other hand, if $\epsilon$ is small, then the
condensate oscillates several times and causes an explosive production
of gamma rays with energy of a few MeV.

\begin{figure}[t]
\centering
\hspace*{-4mm}
\leavevmode\epsfysize=5.6cm \epsfbox{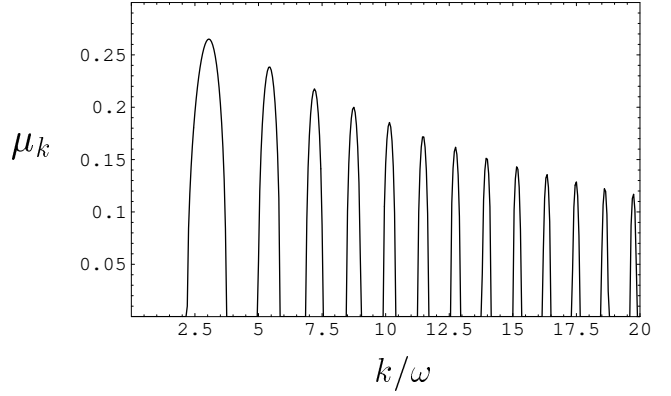}\\[3mm]
\caption[fig3]{\label{fig3} The growth parameters $\mu_k$ as
a function of momenta in units of the effective frequency,
$k/\omega$, after one oscillation of the condensate. 
}
\end{figure}

At density $\rho \approx 10^{14}$~g/cm$^3$ and magnetic field $B \sim
10^{15}$~G, parametric amplification of the MeV photons can take place.
In this case, $\epsilon=0.31$ and $4q_0 \simeq 340$.  We have plotted
$\mu_k$ as a function of $k$, after one oscillation of the condensate, in
Fig.~\ref{fig3}, from which we can deduce the spectrum
\begin{equation}
n_k \simeq {1\over2}\exp(\mu_k\omega t)\,.
\end{equation}
Nuclear matter is opaque for photons with energy of order the electron
chemical potential, which limits the spectral band in which the modes are
amplified.  The spectrum $n_k$ has, therefore, an ultraviolet cut-off at
the momentum associated with the electron Fermi energy inside the neutron
star, beyond which photons are absorbed and the resonance shuts down. This
occurs at wavelengths $k_c \simeq 100 m_e = 51$~MeV. For the model
discussed above, with frequency of oscillations $\omega=4.7$~MeV, this
corresponds to $k/\omega \simeq 11$.

Furthermore, after a few oscillations, backreaction occurs, that is, the
number of created photons is so large that it dominates the frequency of
oscillations of the proton condensate, $m^2 = \omega^2 + 8e^2
\langle\chi^2\rangle$. The backreaction sets in when~\cite{KLS}
\begin{equation}
\langle\langle\chi^2\rangle\rangle = {1\over2\pi^2}\,\int_0^{k_c} dk\,k^2\,
{n_k(t)\over\omega_k} \simeq {\omega^2\over8e^2} \sim \omega^2\,,
\end{equation}
where $\omega_k^2= k^2 + 8e^2\phi^2(t)$ is the effective frequency of
the mode $k$, and the integration is up to the physical cut-off, $k_c$. For
the model in hand, we find that backreaction takes place after about one
oscillation, much before the friction term in Eq.~(\ref{eqm}) has
significantly decreased the oscillation amplitude of the condensate.

\begin{figure}[t]
\centering
\hspace*{-4mm}
\leavevmode\epsfysize=5.4cm \epsfbox{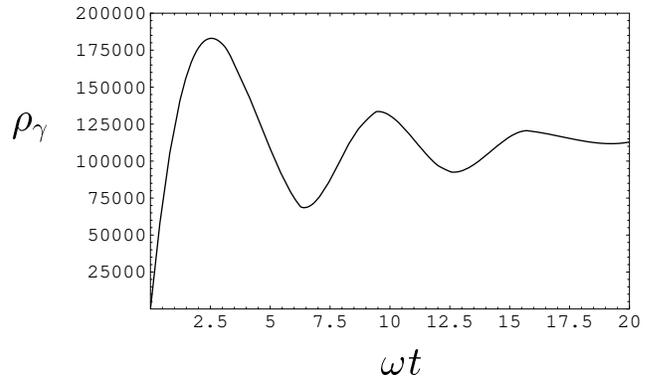}\\[3mm]
\caption[fig4]{\label{fig4} The energy density, in units of MeV$^4$, 
of gamma ray photons produced via parametric resonance, as a function 
of time.}
\end{figure}

The total energy density in photons produced during the resonance is, 
therefore,  
\begin{equation}
\rho_\gamma(t) = {1\over2\pi^2}\,\int_0^{k_c} dk\,k^2\,\omega_k\,
n_k(t)\,,
\end{equation}
We have plotted this energy density in Fig.~\ref{fig4}.
$\rho_\gamma (t)$ reaches an asymptotic value after a few oscillations.
Because of backreaction, the energy density produced is that after the
first oscillation, of order $10^5~{\rm MeV}^4$.  It may seem like a
large value, but actually the energy density produced via this mechanism
is just a small fraction of the neutron
star density, $\rho_\gamma \simeq 8\times 10^4~{\rm MeV}^4 \simeq
10^{-4}\rho $.

We have analyzed the resonant photoproduction numerically for magnetic
field from $10^{12}$~G to $10^{16}$~G and found no significant deviation
from the sample values described above.  The main effect of the magnetic
field is to produce a potential barrier at $\phi=\phi_1$, as shown in
Fig.~\ref{fig1}. A large magnetic field may, however, have some other
effects on nuclear matter.  For example, it can modify the particle
composition, in particular the distribution of protons and
electrons~\cite{chakrabarty}.  This, in turn, can affect the resonant
photoproduction.  Other parameters, such as the electron chemical
potential, the proton fraction $Y_p$, {\it etc.}, can differ significantly
from the sample values we took.  However, our numerical analyses show that
a resonant production of gamma rays from the oscillating charged condensate
is possible for a wide variety of parameters.

\section{Implications for the observation of gamma-ray bursts}

The MeV-energy gamma rays produced in each cycle of oscillations cannot
decay the way they would decay in vacuum, through electron-positron pair
production, because the electrons are highly degenerate inside the
neutron star. Pauli exclusion principle prevents a production of
electrons with energies less than the electron chemical potential, which
is of order 100~MeV.  Compton scattering off the electrons and protons
near the Fermi surface is kinematicly suppressed but is not forbidden.
The corresponding mean free path is $\lambda\sim 10^{-9}$~cm.
Comptonization generally preserves the number of photons and tends to
equalize the temperatures of electrons and photons~\cite{st}.  Given
enough time, the photons would diffuse out of the star.  However, since
the merger of the two neutron stars is characterized by very short time
scales, all the gamma rays non-thermally produced by the coherent
oscillations of the proton condensate remain inside the star when the
fireball erupts.  At that point they can leak out and be observed as a
short gamma-ray burst, or more likely, as a component of the GRB.

The energy stored in the non-thermal bath of gamma ray photons inside
the neutron star is $E_\gamma = \rho_\gamma V \simeq 3\times 10^{50}$~erg.
If the collision of neutron stars releases these photons within a
time scale of order $1~{\rm ms} \lesssim \tau \lesssim 0.1$~s, then the
power generated is of order $10^{52\pm1}$~erg/s, which corresponds 
to the power emitted in the observed gamma ray bursts~\cite{Rees}.

The resonant photoproduction is active in a spherical shell with density
$10^{14} -10^{15} $~g/cm$^3$ that contains most of the neutron star
mass~\cite{st}, and where the acoustically driven superconducting phase
transitions can take place.

\section{Conclusion}

Strong mechanical shock waves, such as those expected to be generated by a
collision of two neutron stars, can cause repetitious superconducting
phase transitions in nuclear matter.  The relaxation of the proton
condensate to its potential minimum can result in the non-thermal resonant
production of $\gamma$-quanta in the MeV energy range.

There are some  important consequences.  The presence of energetic gamma
quanta in nuclear matter, relatively transparent to photons thanks to
electron degeneracy, can affect the equation of state.  This, in turn, can
affect the dynamics of the neutron star coalescence.

In addition, the gamma rays trapped until the onset of the fireball but
then released by the explosion, can also contribute to gamma-ray bursts.

\section*{Acknowledgements}

J.G.B. is supported by a Research Fellowship of the Royal Society.
A. K. is supported in part by the US Department of Energy grant
DE-FG03-91ER40662.


\begin{thebibliography}{99}

\bibitem{Rees} M.~Rees, in Proceedings of Symposium on Black Holes and
  Relativistic Stars, Chicago, IL, 14-15 
  Dec 1996, astro-ph/9701162 (unpublished); P.~M\'esz\'aros, in Proceedings
  of 4th Huntsville Symposium, astro-ph/9711354 (unpublished).

\bibitem{afterglow} J.~van~Paradijs {\it et al.},  Nature {\bf 386}, 686
  (1997); S.~G.~Djorgovski {\it et al.}, Nature {\bf 387}, 876 (1997);
  M.~R.~Metzger {\it et al.}, Nature {\bf 387}, 878 (1997). 

\bibitem{cns} M.~Ruffert, H.-Th.~Janka, and G.~Sch\"afer,
  Astron. Astrophys. 311, 532 (1996); M.~Ruffert, H.-Th.~Janka,
  K.~Takahashi, and G.~Sch\"afer, {\it ibid.} 319, 122 (1997). 

\bibitem{ak} A.~Kusenko, CERN-TH/98-125, astro-ph/9804134; the treatment of
parametric resonance in this paper contained some errors. 

\bibitem{super} G.~Baym, C.~Pethick, and D.~Pines, Nature {\bf 224}, 673
  (1969);  G.~Baym, C.~Pethick, D.~Pines, and M.~Ruderman, Nature {\bf
    224}, 872 (1969).   

\bibitem{egap} N.-C.~Chao, J.~W.~Clarck, and C.-H.~Yang, Nucl. Phys. {\bf
    A179}, 320 (1972). 

\bibitem{chen} J.~M.~C.~Chen, J.~W.~Clarck, R.~D.~Dav\'e, and V. V. Khodel,
Nucl. Phys. {\bf A555}, 59 (1993). 

\bibitem{spectrum} B.~W.Carroll, E.~G.~Zweibel, C.~J.~Hansen,
  P.~N.~McDermott, M.~P.~Savedoff, J.~H.~Thomas, and H.~M.~Van~Horn,
  Astrophys. J. {\bf 305}, 767 (1986); R.~I.~Epstein, Astrophys. J. {\bf
  333}, 880 (1988). 

\bibitem{param} A. A. Grib, S. G. Mamayev, and V. M. Mostepanenko, {\it
    Vacuum quantum effects in strong fields},  Fridmann Laboratory,
    St. Petersburg, 1994.   

\bibitem{tinkham} M.~Tinkham, {\it Introduction to superconductivity},
  McGraw Hill, New York, 1996.   

\bibitem{at} E.~Abrahams and T.~Tsuneto, Phys. Rev. {\bf 152}, 416 (1966).

\bibitem{ge} L.~P.~Gor'kov and G.~M.~Eliashberg, JETP 27, 328 (1968).
  [Zh. Eksp. Teor. Fiz. 54, 612 (1968)]. 

\bibitem{ll} L.~D.~Landau and E.~M.~Lifshitz,  {\it Course of
    theoretical physics}, vol. 9, Pergamon Press, New York, 1980. 

\bibitem{tkachev} I.~I.~Tkachev, Sov. Astron. Lett. {\bf 12} 305 (1986);
Phys. Lett. {\bf B 191} 41 (1987). 

\bibitem{KLS} L. Kofman, A. D. Linde, and A. A.~Starobinsky,
 Phys. Rev. Lett.  {\bf 73}, 3195 (1994); 
 Phys. Rev. {\bf D 56}, 3258 (1997).

\bibitem{GBL} J. Garc\'\i a-Bellido and A.D. Linde,
 Phys. Rev. {\bf D 57}, 6075 (1998).

\bibitem{math} N. W. McLachlan, {\it Theory and application of Mathieu
functions}, Dover, New York, 1964.

\bibitem{chakrabarty} S.~Chakrabarty, D. Bandyopadhyay, and S.~Pal,
Phys. Rev. Lett. {\bf 78}, 2898 (1997). 

\bibitem{st} S.~L.~Shapiro and S.~A.~Teukolsky, {\it Black Holes, White
  Dwarfs, and Neutron Stars}, John Wiley \& Sons, New York, 1986.

\end{thebibliography}
\end{document}